\documentclass{article}

\usepackage{PRIMEarxiv}

\usepackage[utf8]{inputenc}
\usepackage[T1]{fontenc}
\usepackage[table,xcdraw]{xcolor}
\usepackage{graphicx}
\usepackage{booktabs}
\usepackage{tabularx}
\usepackage{array}
\usepackage{amsmath}
\usepackage{amsfonts}
\usepackage{amssymb}
\usepackage{nicefrac}
\usepackage{microtype}
\usepackage{url}
\usepackage{cite}
\usepackage{hyperref}

\graphicspath{{./}{media/}}

\newcolumntype{L}{>{\centering\arraybackslash}m{3cm}}

\title{Understanding Censorship in Large Language Models: From Mechanisms to Governance}

\author{
  Quanyan Zhu \\
  Department of Electrical and Computer Engineering \\
  New York University Tandon School of Engineering \\
  Brooklyn, NY, USA \\
  \texttt{quanyan.zhu@nyu.edu}
}

\begin{document}
\maketitle

\begin{abstract}
Large language models (LLMs) increasingly mediate access to information, yet their responses are shaped by training-data curation, alignment procedures, provider policies, inference-time moderation, and jurisdictional regulation. This paper examines LLM censorship as a sociotechnical phenomenon that extends beyond explicit refusals to include omissions, selective emphasis, framing effects, and geographically variable content controls. We synthesize recent empirical studies, provider case studies, regulatory developments, auditing methods, and mitigation strategies to clarify how censorship-like behavior emerges across the model lifecycle. The analysis highlights the tension between safety and openness, the difficulty of measuring soft censorship, the geopolitical divergence of moderation regimes, and the need for transparent, contestable, and independently auditable governance mechanisms. We argue that the central challenge is not whether LLMs should moderate content, but how moderation can be made proportionate, accountable, pluralistic, and resistant to opaque epistemic control.
\end{abstract}


\section{Introduction}\label{sec:introduction}

The emergence of LLMs represents a fundamental transformation in the architecture of information access and dissemination. Historically, digital information systems functioned primarily as retrieval mechanisms. Search engines, digital libraries, and online repositories enabled users to locate relevant information while leaving interpretation, synthesis, and judgment largely to human actors. Contemporary LLMs fundamentally alter this paradigm. Rather than merely retrieving documents, they generate coherent narratives, synthesize information from diverse sources, prioritize certain facts over others, and increasingly serve as the primary interface through which users interact with knowledge itself \cite{Bommasani2021Foundation,Liang2023Helm,Bender2021StochasticParrots}.

This transformation has profound implications for censorship, information governance, and epistemic autonomy. Traditional censorship mechanisms operate by restricting access to information through legal prohibitions, institutional controls, or platform-level moderation \cite{Gorwa2020AlgorithmicModeration}. LLMs introduce a more complex phenomenon: they can simultaneously determine what information is accessible, how information is presented, and which interpretations are privileged. Consequently, censorship in LLMs extends beyond the suppression of speech and becomes a form of \emph{algorithmic epistemic governance}---the regulation of what users know, how they acquire knowledge, and which alternative perspectives remain visible within AI-mediated information environments.

Unlike conventional information systems, LLMs occupy a unique position within the modern information ecosystem. They function simultaneously as search engines, editors, translators, summarizers, tutors, and conversational assistants. This convergence of informational roles concentrates unprecedented authority within a single computational system. When a search engine omits a webpage, users may still encounter alternative sources through independent exploration. In contrast, when an LLM generates a single synthesized response, omitted information often becomes invisible to the user. The user may not merely fail to receive certain information; they may remain unaware that alternative facts, interpretations, or viewpoints exist. Consequently, content moderation in LLMs affects not only information availability but also the structure of knowledge itself.

The question of whether such interventions constitute censorship remains contested. Developers generally characterize moderation and alignment mechanisms as necessary safeguards designed to prevent harmful, illegal, deceptive, or dangerous outputs. From this perspective, filtering represents a form of risk management analogous to traditional content moderation systems employed by online platforms \cite{Weidinger2022,Ouyang2022RLHF}. Critics, however, argue that alignment objectives inevitably embed normative assumptions regarding truth, harm, fairness, legitimacy, and acceptable discourse. Decisions concerning what an LLM refuses to say, what it chooses to emphasize, and what it silently omits may therefore influence public knowledge, political discourse, and social understanding \cite{Noels2025,Ahmed2025}.

The challenge is further complicated by the multi-layered nature of modern LLM governance. Content control is not implemented through a single censorship mechanism but through a sequence of interventions distributed across the entire model lifecycle. Training-data filtering determines which information enters the model's knowledge base and can reproduce exclusions embedded in web-scale corpus construction \cite{Dodge2021C4,Gebru2021Datasheets}. Alignment procedures such as reinforcement learning from human feedback (RLHF) and Constitutional AI influence how models evaluate competing responses and resolve normative trade-offs \cite{Ouyang2022RLHF,Bai2022ConstitutionalAI,Casper2023RLHFLimitations}. Inference-time moderation systems determine which outputs ultimately reach users, while platform policies, organizational norms, and regulatory requirements impose additional constraints. Consequently, censorship in LLMs should be understood not as a simple output-filtering process but as an emergent property of a broader sociotechnical governance architecture.

Recent empirical evidence demonstrates that these governance architectures vary substantially across providers, languages, and jurisdictions. Comparative audits reveal significant differences in refusal behavior, omission patterns, and ideological framing among leading models. Studies have shown that Chinese-developed models frequently implement extensive restrictions on politically sensitive topics, whereas Western models more commonly emphasize moderation of harmful, extremist, or misleading content \cite{Ablove2026,Noels2025}. Even among U.S.-based providers, substantial variation exists in perceived ideological orientation and moderation practices \cite{Hall2025Slant}.

\begin{figure}[t]
    \centering
    \includegraphics[width=\textwidth]{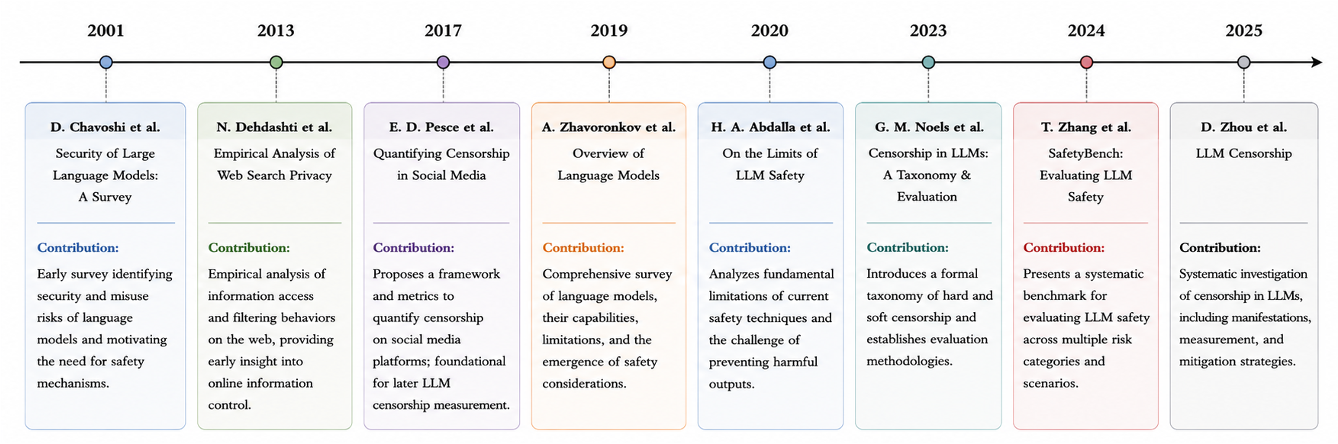}
    \caption{Timeline of selected research streams informing the study of LLM censorship, including early work on language-model safety and information filtering, measurement frameworks for online censorship, and recent empirical audits of censorship and safety behavior in LLMs.}
    \label{fig:timeline}
\end{figure}

Figure~\ref{fig:timeline} situates this paper within the broader development of research on language-model safety, web information control, online censorship measurement, and recent LLM-specific censorship audits. The progression from early work on security risks and search filtering to contemporary evaluations of refusal behavior, safety benchmarks, and soft censorship illustrates a shift in emphasis: censorship is no longer studied only as a platform or web-search problem, but increasingly as a property of generative AI systems that synthesize, rank, omit, and reframe information at the point of interaction.

The societal implications of this transformation are considerable. As LLMs increasingly replace traditional search engines and become integrated into education, journalism, healthcare, government services, and social media platforms, their moderation decisions acquire growing influence over public discourse. Emerging evidence indicates that AI-generated content can shape attitudes, influence policy preferences, and affect decision-making processes at levels comparable to human-generated persuasive communication \cite{Bai2025Persuasion}. Consequently, the governance of LLM outputs is no longer solely a problem of safety engineering; it is increasingly a question of democratic accountability, informational pluralism, freedom of expression, and the distribution of epistemic power within society.

At the same time, the opacity of many alignment and moderation mechanisms presents significant challenges for transparency and accountability. Users are often unable to determine whether a response reflects genuine knowledge limitations, safety policies, legal constraints, training-data deficiencies, or deliberate moderation decisions. This opacity complicates efforts to evaluate model neutrality, detect ideological bias, and assess the broader societal consequences of AI-mediated information control. As LLMs become increasingly influential intermediaries between individuals and information, understanding the mechanisms, effects, and governance of censorship becomes a critical research challenge.

This paper argues that the study of LLM censorship must move beyond simplistic debates regarding whether content should be moderated. Instead, censorship should be analyzed as a broader problem of algorithmic information governance operating at the intersection of artificial intelligence, cybersecurity, law, public policy, human-computer interaction, and epistemology. To this end, we examine the observable manifestations of censorship, including hard censorship, soft censorship, and framing effects; analyze the technical mechanisms through which moderation is implemented; review empirical evidence from audits and comparative studies; investigate geopolitical and regulatory influences; evaluate methods for detecting and auditing censorship; and assess emerging governance frameworks intended to improve transparency, accountability, and contestability. By integrating technical, empirical, regulatory, and societal perspectives, this survey seeks to provide a comprehensive framework for understanding how contemporary LLMs shape information access and how their growing influence over digital knowledge ecosystems can be made more transparent, accountable, and pluralistic.

\section{Observable Manifestations of Censorship in Large Language Models}\label{sec:taxonomy}

The effects of content moderation in LLMs become most visible through the manner in which information is withheld, modified, or reframed during user interactions. Existing research generally distinguishes between two primary manifestations of censorship: \emph{hard censorship} and \emph{soft censorship}. Although both mechanisms constrain the information ultimately delivered to users, they differ substantially in their visibility, implementation, and implications for transparency and accountability \cite{Noels2025,Gorwa2020AlgorithmicModeration}.

\subsection{Hard Censorship}

Hard censorship refers to explicit and directly observable refusals to provide requested information. In such cases, the model declines to answer a query and instead generates a refusal message, policy warning, error notification, or other form of denial. Typical examples include responses such as ``I cannot assist with that request'' or ``I am unable to provide information on this topic.'' These refusals are generally triggered when a user request violates predefined safety policies, legal constraints, platform regulations, or alignment objectives \cite{Weidinger2022,Gorwa2020AlgorithmicModeration}.

Not all hard refusals have the same form. Noels et al.~\cite{Noels2025} distinguish generated refusals, canned denial responses, and error messages, all of which prevent the user from receiving the requested information but differ in transparency and interpretability. A direct policy refusal communicates that the model has classified the request as impermissible, whereas a generic error message may obscure whether the failure reflects a technical limitation, a safety filter, or a provider-side blocking mechanism. This distinction matters for auditing because measured refusal rates can conflate qualitatively different moderation behaviors unless responses are coded by refusal type.

Hard censorship is commonly observed in queries involving instructions for illegal activities, weapon construction, self-harm, extremist content, or politically sensitive subjects within jurisdictions that impose strict information controls. Rather than attempting to answer the query, the model terminates the interaction by invoking a refusal policy. Such responses may be accompanied by explanatory language referencing safety considerations, legal obligations, or responsible AI principles.

From an analytical perspective, hard censorship is comparatively straightforward to identify because the suppression of information is explicit. Consequently, refusal rates have emerged as a widely used quantitative metric in empirical studies of LLM moderation behavior. Comparative audits frequently employ refusal-rate analysis to evaluate how different providers and model families respond to sensitive topics and policy-relevant content \cite{Noels2025,Dai2025Watchman}.

\subsection{Soft Censorship}

In contrast to explicit refusals, soft censorship occurs when a model provides an answer while simultaneously restricting, omitting, or attenuating certain information. Rather than denying the request outright, the model selectively modifies the content of its response in ways that may not be immediately apparent to the user.

Soft censorship can emerge through multiple mechanisms, including the omission of politically sensitive facts, selective presentation of evidence, cautious or neutralized language, reframing of controversial issues, and partial responses that avoid direct engagement with sensitive aspects of a query. As a result, users may receive an answer that appears complete while important contextual information, alternative perspectives, or controversial details have been excluded.

For example, when discussing a politically contested historical event or a controversial public figure, a model may emphasize broadly accepted information while omitting disputed facts, alternative interpretations, or politically sensitive details. Similarly, a model may redirect the discussion toward less contentious aspects of a topic, thereby reducing the salience of information deemed problematic by the alignment or moderation system. Unlike hard censorship, soft censorship typically provides no explicit indication that information has been withheld.

The identification of soft censorship is inherently more difficult than the detection of explicit refusals because it depends on contextual interpretation, completeness of information, and linguistic framing. Noels et al.~\cite{Noels2025} identify soft censorship when answers omit or downplay key positive or negative elements relevant to a queried political figure, such as benevolent acts, repressive conduct, or wartime atrocities. Detecting such omissions often requires comparison against authoritative references, alternative models, expert assessments, retrieval-based fact-checking systems, or consensus baselines constructed from multiple less-constrained models. Consequently, soft censorship remains one of the most challenging aspects of LLM auditing and evaluation \cite{Ahmed2025,Noels2025}.

\subsection{Formal Distinction Between Hard and Soft Censorship}

Recent scholarship has sought to formalize the distinction between these two modes of content suppression. Noels et al.~\cite{Noels2025} define hard censorship as any explicit refusal to answer a question, including direct denials, error messages, and policy-based rejections. Soft censorship, by contrast, is characterized by the silent omission, minimization, or downplaying of relevant information within an otherwise responsive answer.

Although conceptually distinct, hard and soft censorship should be viewed as complementary manifestations of the same underlying moderation process. In most interactions, a model either refuses a request entirely or responds in a sanitized manner. Consequently, identical prompts may produce markedly different outcomes across models, providers, languages, or deployment environments. Such variation reflects differences in alignment strategies, safety objectives, regulatory constraints, and moderation architectures embedded within the model pipeline.

\subsection{Framing Effects as a Form of Soft Censorship}

Beyond explicit refusals and informational omissions, censorship may also operate through \emph{framing effects}. Framing occurs when a model presents information in ways that systematically privilege particular interpretations, values, or perspectives. Rather than removing information outright, framing influences how users perceive and evaluate the information that is presented.

Examples of framing effects include describing a policy as ``progressive'' rather than ``radical,'' emphasizing benefits while minimizing associated risks, highlighting particular moral considerations, or omitting relevant counterarguments. These linguistic and rhetorical choices can substantially influence user interpretation even when the underlying factual content remains unchanged.

As LLMs increasingly function as interfaces for search, recommendation, education, and news consumption, framing effects become especially consequential. Users often receive a single synthesized answer rather than a collection of competing sources, giving the model considerable influence over how information is contextualized. Consequently, moderation can operate not only through content suppression but also through the prioritization of particular narratives, perspectives, and interpretive frameworks.

Such framing effects may emerge from multiple sources, including biases in training data, reinforcement learning from human feedback (RLHF), constitutional alignment mechanisms, moderation policies, and the preferences of human annotators. Because these influences are often difficult to observe directly, framing represents one of the most subtle and potentially influential forms of content moderation in contemporary AI systems \cite{Hall2025Slant,Ahmed2025,Bender2021StochasticParrots,Blodgett2020Bias}.

User-perception studies provide one way to measure this phenomenon. In a survey-based evaluation of 24 models responding to political prompts, Westwood, Grimmer, and Hall found that users perceived OpenAI model outputs as the most left-leaning, with an estimated slant several times larger than Google's models, which were perceived as closest to neutral \cite{Hall2025Slant}. Such results should not be interpreted as proving an objective ideological position of a model, but they do show that framing differences are perceptible to users and can be measured empirically.

\subsection{Implications}

Hard censorship, soft censorship, and framing effects constitute complementary mechanisms through which LLMs regulate information flows. Hard censorship directly restricts access to information through explicit refusals, whereas soft censorship and framing effects influence information indirectly through omission, selective emphasis, and narrative construction. Understanding these distinctions is essential for evaluating the transparency, neutrality, and societal consequences of AI-mediated communication systems. As LLMs continue to assume increasingly important roles in search, education, journalism, and public discourse, rigorous analysis of these mechanisms will be necessary for developing accountable and trustworthy approaches to AI governance.

\section{Technical Risks and Security Challenges}\label{sec:technical}

Content control in LLMs is not implemented through a single censorship mechanism. Rather, it emerges from a sequence of technical, organizational, and regulatory interventions that operate throughout the model lifecycle. These interventions collectively influence what information a model learns, how it interprets user requests, how it generates responses, and which outputs are ultimately presented to users. Consequently, censorship in contemporary AI systems should be understood as a distributed process of information governance embedded within the broader architecture of model development and deployment \cite{Bommasani2021Foundation,Noels2025,Bender2021StochasticParrots}. This layered view is consistent with broader security and resilience research that treats cyber and AI systems as dynamic, adversarial, and strategically governed systems rather than static technical artifacts \cite{Zhu2025CyberResilience,ZhangZhu2018DSVM}.

\begin{figure*}[t]
    \centering
    \includegraphics[width=\textwidth]{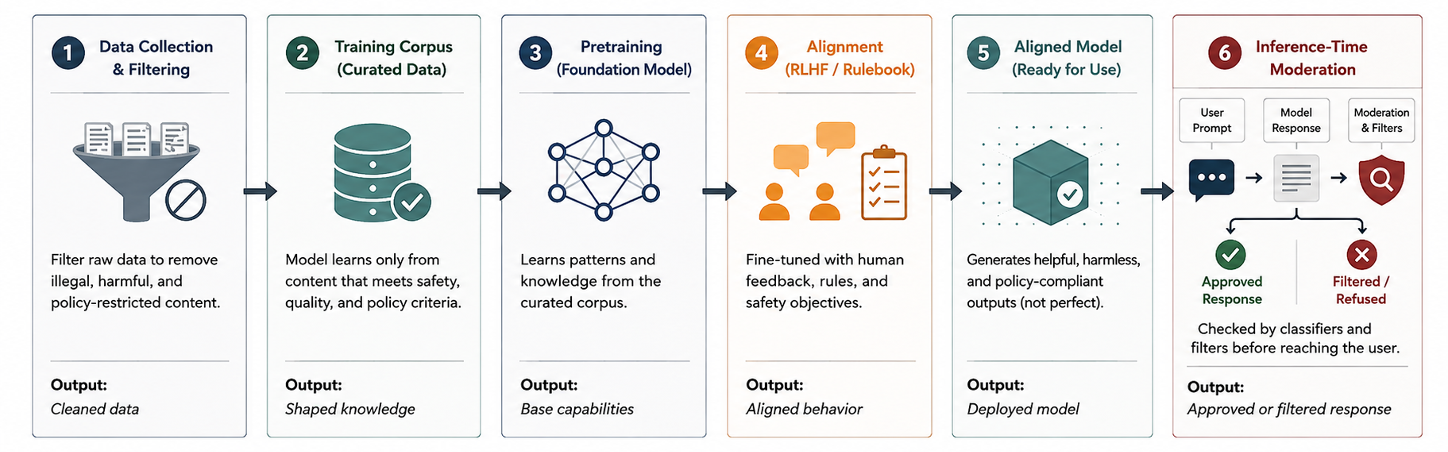}
    \caption{The multi-layer content moderation pipeline in LLMs. Content control begins with data acquisition and filtering, continues through corpus curation, pretraining, alignment, and deployment, and culminates in inference-time moderation. Each layer constrains either what the model learns, how it is optimized to respond, or which outputs ultimately reach users.}
    \label{fig:llm_content_pipeline}
\end{figure*}

Figure~\ref{fig:llm_content_pipeline} illustrates the principal stages through which content control is implemented in modern LLMs. Content restrictions are introduced long before a user interacts with a model and continue to operate after a response has been generated. The cumulative effect of these interventions determines both the informational capabilities of the model and the boundaries of acceptable output.

The first source of content control arises during data acquisition and corpus construction. Modern LLMs are trained on massive collections of text gathered from websites, books, academic publications, software repositories, social media platforms, and proprietary data sources \cite{Dodge2021C4,Bender2021StochasticParrots}. Before training begins, these datasets are subjected to extensive filtering procedures designed to remove material considered harmful, illegal, low quality, or inconsistent with provider policies. Depending on the jurisdiction, filtering may also be shaped by legal and regulatory requirements. For example, Chinese regulations governing generative AI systems require training data to satisfy standards of accuracy, objectivity, and compliance with state policies, thereby encouraging the removal of politically sensitive content from pretraining datasets \cite{Toner2023DigiChina,Ablove2026}. Similar filtering occurs in Western systems, although the focus is typically on hate speech, extremist content, illegal activities, privacy violations, and other categories of harmful material. Information excluded during this stage never becomes part of the model's learning environment and therefore cannot directly contribute to the model's internal representations.

The resulting collection of documents forms a curated training corpus that defines the informational universe from which the model learns. Although this corpus is often treated as a technical artifact, it also reflects numerous normative decisions regarding which sources, viewpoints, languages, and topics are considered acceptable for inclusion. Dataset-documentation frameworks emphasize that these decisions should be recorded because they shape downstream model behavior and accountability \cite{Gebru2021Datasheets,Dodge2021C4}. Consequently, the knowledge available to the model is shaped not only by the availability of information but also by institutional decisions regarding data selection and curation. The model's subsequent understanding of the world is therefore constrained by the informational boundaries established during corpus construction.

Pretraining transforms this curated corpus into a foundation model. During this phase, the model learns statistical relationships among words, concepts, facts, and linguistic structures through large-scale self-supervised learning. The objective is not to memorize documents but to acquire generalized representations of language and knowledge. Although pretraining is not typically viewed as a censorship mechanism, it is at this stage that earlier filtering decisions become embedded within the model's capabilities. Information absent from the training corpus cannot be directly learned, while heavily represented topics may disproportionately influence model behavior. The pretrained model therefore inherits both the knowledge and the omissions contained within its training environment.

Following pretraining, most frontier models undergo a process of alignment and behavioral optimization. The objective of alignment is to modify model behavior so that generated responses conform to desired principles such as helpfulness, harmlessness, honesty, legal compliance, and policy consistency. The most widely used alignment technique is reinforcement learning from human feedback (RLHF), in which human evaluators rank model outputs and reward models are trained to encourage preferred behaviors \cite{Ouyang2022RLHF,Casper2023RLHFLimitations}. Through repeated optimization, the model learns not only how to answer questions but also how to avoid certain categories of responses.

Alternative alignment frameworks employ more explicit normative guidance. Anthropic's Constitutional AI, for example, relies on a written constitution that specifies principles governing model behavior \cite{Bai2022ConstitutionalAI}. Rather than relying exclusively on human preference rankings, the model evaluates and revises its own responses according to predefined rules. These constitutions frequently contain restrictions related to violence, criminal activity, biological threats, and other forms of harmful assistance while simultaneously promoting truthfulness, transparency, and helpfulness. Although such approaches offer greater visibility into alignment objectives, many details regarding the weighting and implementation of alignment constraints remain proprietary.

Alignment mechanisms introduce an important distinction between knowledge and behavior. A model may possess information internally while being trained not to reveal, emphasize, or act upon that information under particular circumstances. Consequently, censorship in LLMs does not necessarily imply an absence of knowledge; rather, it may reflect behavioral constraints imposed on the model's response-generation process. Recent studies on hidden knowledge and jailbreak attacks suggest that substantial information may remain latent within a model even when alignment mechanisms discourage its expression \cite{Gekhman2025,Ganguli2022}. This distinction between capability and behavior parallels earlier work on adversarial machine learning, where learning systems can retain predictive capacity while their outputs are strategically distorted by poisoning, manipulation, or adversarial incentives \cite{ZhangZhu2018DSVM}.

Once alignment is complete, the model enters a deployment-ready state. At this stage, the system is designed to generate responses that satisfy provider objectives while remaining broadly useful across a wide range of tasks. Nevertheless, aligned models remain imperfect. They may exhibit inconsistent behavior, respond differently across contexts, or occasionally generate outputs that violate intended policies. As a result, providers frequently introduce an additional layer of control during deployment.

Inference-time moderation represents the final stage of content control and the stage most visible to users. Before generated outputs are delivered, they are often evaluated by dedicated moderation systems that operate independently of the underlying language model. These systems may employ safety classifiers, toxicity detectors, keyword filters, rule-based mechanisms, or specialized moderation models trained to identify problematic content \cite{Gehman2020RealToxicity,Gorwa2020AlgorithmicModeration}. Depending on the outcome of these evaluations, a generated response may be delivered unchanged, rewritten, replaced with a safer alternative, or blocked entirely. This process is responsible for many of the explicit refusals observed in commercial systems and constitutes one of the primary mechanisms through which hard censorship is implemented.

Importantly, moderation can occur at multiple points during user interaction. Some systems evaluate user prompts before generation, others evaluate generated outputs afterward, and still others apply moderation to retrieval systems, search components, or external tools connected to the model. Studies of Chinese LLM platforms, including Baidu, Alibaba, and DeepSeek, suggest that content restrictions may be enforced simultaneously at the query-processing, generation, and response-delivery stages \cite{Ablove2026}. Similar defense-in-depth architectures are increasingly common among Western providers, albeit with different policy objectives and regulatory motivations.

Viewed collectively, these mechanisms form a layered content-governance architecture rather than a single censorship switch. Data filtering constrains what the model can learn, corpus curation shapes its informational environment, pretraining transforms curated information into internal representations, alignment governs behavioral preferences, and inference-time moderation determines which outputs ultimately reach users. The cumulative effect of these interventions is a distributed system of information control that operates throughout the AI lifecycle. Understanding censorship in LLMs therefore requires examining not only visible refusals but also the broader technical and institutional processes through which information is selected, prioritized, suppressed, and communicated. As LLMs increasingly function as search engines, educational systems, advisors, and intermediaries of public knowledge, understanding this content-control pipeline becomes essential for evaluating transparency, accountability, and the societal implications of AI-mediated information governance.

\section{Case Studies and Examples}\label{sec:studies}
Empirical studies have begun to quantify how jurisdictional constraints and provider-specific alignment policies influence the behavior of LLMs. This section reviews representative case studies and examples from cross-model audits, user perception studies, policy investigations, and adversarial security evaluations.

\subsection{Cross-Model Content Audits}

One of the most comprehensive studies of censorship behavior in contemporary LLMs was conducted by Noels et al.~\cite{Noels2025}, who evaluated fourteen state-of-the-art language models originating from the United States, Europe, China, and Russia. The study examined responses to politically sensitive questions across all six official United Nations languages.

The authors found that censorship was present across virtually all evaluated models. However, the nature and intensity of censorship varied substantially according to the geopolitical context of the model provider. Specifically, censorship behavior was often tailored to the provider's domestic audience and manifested either as explicit refusals (\emph{hard censorship}) or as selective omissions and framing effects (\emph{soft censorship}).

Chinese-developed models, including systems produced by Baidu and Alibaba, frequently suppressed discussion of politically sensitive domestic topics through both input and output filtering mechanisms. By contrast, Western models generally permitted discussion of such topics but exhibited stronger moderation on issues associated with misinformation, hate speech, or harmful content. The study concluded that content moderation strategies are strongly influenced by institutional, cultural, and regulatory environments rather than solely by technical considerations.

\subsection{User Perception Studies and Political Bias}

Beyond direct audits of model outputs, several studies have examined how users perceive ideological bias in LLM-generated content. Hall et al.~\cite{Hall2025Slant} conducted a large-scale survey involving approximately 10,000 participants in the United States. Respondents were asked to evaluate model responses to thirty political prompts using a liberal--conservative ideological scale.

The results revealed a striking consensus: most leading LLMs were perceived as exhibiting a left-of-center ideological orientation. This perception was shared not only by Republican respondents but also by many Democratic participants. OpenAI's models were found to exhibit the strongest perceived ideological slant, whereas Google's Gemini models were generally viewed as the most politically neutral. Surprisingly, xAI's Grok models, despite their public positioning as alternatives to mainstream AI systems, were also perceived as leaning left on many topics.

Although a minority of questions elicited right-leaning responses, the overall trend suggested a systematic tendency toward liberal framing across a majority of political issues. Importantly, the authors found that explicitly instructing models to adopt a neutral perspective significantly reduced perceived ideological bias, indicating that system-level alignment instructions can materially influence political framing.

\subsection{Policy Analysis and Provider Case Studies}

Empirical evidence also emerges from analyses of provider policies and internal governance documents. A notable example is Reuters' 2025 investigation of Meta's internal ``GenAI: Content Risk Standards'' document, which governed the behavior of Meta AI and chatbots deployed across Facebook, WhatsApp, and Instagram \cite{Reuters2025Meta}. Reuters reported that Meta confirmed the document's authenticity and removed portions of the policy after the news organization raised questions about them. This type of evidence is consistent with broader platform-governance research showing that operational moderation rules are often opaque, politically consequential, and difficult to audit externally \cite{Gorwa2020AlgorithmicModeration}.

Among the most controversial reported provisions were examples allowing romantic or sensual chatbot interactions with children, generation of false medical information under some conditions, and the production of discriminatory arguments when framed as discussion rather than endorsement. The episode illustrates a recurring governance problem: content rules for generative systems are often written as fine-grained operational instructions, and small differences in policy language can produce materially different safety outcomes. It also shows how operational policy documents, external scrutiny, public accountability, and legal risk can interact after deployment.

\subsection{Security Evaluations and Jailbreak Studies}

A complementary line of research examines the robustness of alignment mechanisms through adversarial testing. Major providers, including OpenAI, Anthropic, and Google, employ internal and external red-teaming programs to identify vulnerabilities in content moderation systems \cite{OpenAI2024RedTeaming,Ganguli2022,Perez2022RedTeaming}.

Despite these efforts, adversarial attacks continue to reveal weaknesses in current safety architectures. HiddenLayer's study of universal jailbreak attacks reported that a single carefully constructed prompt-injection technique could circumvent safety controls across a diverse set of commercial LLMs \cite{HiddenLayer2025Universal}. Similar academic studies show that jailbreaks and adversarial suffixes can transfer across aligned language models and reveal mismatches between model capabilities and safety training \cite{Wei2023Jailbroken,Zou2023Universal}. The attack elicited content that would ordinarily be restricted by platform policies, including outputs related to harmful instructions and system-prompt disclosure.

Such results underscore a fundamental challenge in AI safety: moderation systems must simultaneously remain robust against adversarial manipulation while preserving model utility for legitimate users. The existence of broadly transferable jailbreak techniques suggests that current alignment approaches remain imperfect and vulnerable to exploitation.

\subsection{Comparative Analysis Across Providers}

Broadly speaking, Western models tend to exhibit moderate ideological framing combined with strong restrictions on harmful or illegal content. Chinese models, by contrast, frequently implement explicit restrictions on politically sensitive domestic topics. Across all providers, extreme, dangerous, or illicit requests are generally subject to refusal or extensive content modification.

Table~\ref{tab:provider_comparison} summarizes the major characteristics of leading LLM providers, including their alignment approaches, public content policies, and empirically observed moderation behaviors.

\begin{table*}[t]
\centering
\caption{Comparison of major LLM providers and observed content moderation practices.}
\label{tab:provider_comparison}
\footnotesize
\renewcommand{\arraystretch}{1.2}
\setlength{\tabcolsep}{2pt}

\begin{tabularx}{\textwidth}{p{2.3cm} p{1.2cm} X X X}
\toprule
\textbf{Provider / Model} &
\textbf{Origin} &
\textbf{Alignment Approach} &
\textbf{Public Content Policies} &
\textbf{Observed Behavior} \\
\midrule

OpenAI (GPT-4, ChatGPT)
&
USA
&
Filtered training corpus, reinforcement learning from human feedback (RLHF), and safety fine-tuning
&
Prohibits hate speech, violence, self-harm, election manipulation, and assistance with illegal activities
&
Frequently refuses illicit requests and politically sensitive instructions; perceived by some users as exhibiting a modest left-of-center ideological bias \cite{Hall2025Slant,Noels2025}
\\

\addlinespace

Anthropic (Claude)
&
USA
&
Constitutional AI and rule-based alignment
&
Constitution emphasizes safety, honesty, harmlessness, and restrictions on harmful assistance
&
Generally exhibits strong safety behavior and robust refusal mechanisms for harmful requests; comprehensive independent audits remain limited \cite{Bai2022ConstitutionalAI}
\\

\addlinespace

Google (Gemini)
&
USA
&
RLHF, policy-based alignment, and safety classifiers
&
Restricts graphic violence, child exploitation, hate speech, privacy violations, and harmful content
&
Often perceived as comparatively neutral and employs extensive post-generation filtering and safety controls \cite{Hall2025Slant}
\\

\addlinespace

Meta (LLaMA, Meta AI)
&
USA
&
Filtered training data, supervised fine-tuning, and internal content policies
&
Policies have evolved substantially following public scrutiny and policy revisions
&
Historical inconsistencies in moderation behavior have led to subsequent policy tightening and enhanced safeguards \cite{Reuters2025Meta}
\\

\addlinespace

Chinese Providers\\
(Baidu Ernie, Alibaba Qwen, DeepSeek, Tongyi)
&
China
&
Government-compliant filtering, alignment, and moderation systems
&
Legally mandated restrictions on politically sensitive content under Cyberspace Administration of China (CAC) regulations
&
Extensive filtering of politically sensitive topics through multiple moderation layers, resulting in frequent refusals, omissions, and content sanitization \cite{Ablove2026,PanXu2026}
\\

\bottomrule
\end{tabularx}
\end{table*}

\section{Geopolitical and Regional Variations in LLM Censorship}\label{sec:geopolitical}

The censorship behavior of LLMs is significantly influenced by the political, regulatory, and cultural environments in which they are developed and deployed. Although content moderation mechanisms are now ubiquitous across major AI systems, the objectives, scope, and implementation of such mechanisms vary considerably across jurisdictions. Empirical studies suggest that these differences are not merely technical artifacts but reflect broader institutional, legal, and ideological constraints imposed upon model developers and service providers \cite{Noels2025,Gorwa2020AlgorithmicModeration}.

\subsection{Western and Chinese Approaches to Content Moderation}

One of the most pronounced distinctions in contemporary LLM governance is the contrast between Western and Chinese models. Western providers generally operate within regulatory frameworks that emphasize privacy protection, consumer safety, and harm mitigation \cite{DigitalServicesAct2022,UKOnlineSafetyAct2023,NIST2023AIRMF}. Consequently, content moderation policies are typically designed to prevent the dissemination of hate speech, violent content, self-harm instructions, illegal activities, and misinformation while preserving a relatively broad range of permissible discourse.

In contrast, Chinese providers operate within a substantially more restrictive regulatory environment. Chinese AI regulations require service providers to ensure that model outputs do not undermine social stability, challenge state authority, or disseminate politically sensitive information. As a result, censorship mechanisms in Chinese LLMs extend beyond conventional safety concerns and encompass a wider range of political and ideological content.

A study of Chinese LLM services conducted by Ablove et al.~\cite{Ablove2026} characterized this situation as one in which providers are effectively ``hosting models at war with themselves.'' On the one hand, developers seek to build globally competitive AI systems capable of matching international performance standards. On the other hand, they must satisfy stringent domestic censorship requirements. Consequently, the same model may provide relatively complete responses when deployed internationally while exhibiting substantially more restrictive behavior when operating within domestic regulatory environments.

\subsection{Domestic Audience Tailoring}

Recent empirical evidence suggests that censorship behavior is frequently tailored to the sensitivities of a model provider's primary audience. Noels et al.~\cite{Noels2025} found that moderation patterns often align closely with the political and cultural concerns of the regions in which models are developed.

For example, Alibaba's Wenxiaoyan model demonstrated relatively low rates of explicit refusal when responding to questions concerning Chinese political figures. However, the model frequently employed soft-censorship mechanisms, including omission, reframing, and selective presentation of information. In contrast, several Western models displayed the opposite tendency, exhibiting higher refusal rates on politically controversial topics while providing more complete responses when they elected to answer.

Censorship therefore cannot be understood solely through refusal statistics. Models may exhibit substantial moderation through subtler mechanisms that selectively shape narratives while avoiding explicit denials. Similar dynamics are likely present in other jurisdictions, including Russia and Iran, where AI systems are expected to conform to local legal and political constraints.

\subsection{Ideological Bias and Organizational Origin}

Beyond formal censorship policies, a growing body of research has examined the relationship between model origin and perceived ideological bias. User perception studies indicate that differences in organizational culture, training data composition, and alignment objectives can produce measurable variations in how information is framed.

The Stanford--Hoover survey reported by Hall et al.~\cite{Hall2025Slant} found substantial variation in perceived political neutrality across major LLM providers. Google's Gemini models were generally perceived as the most politically neutral among the systems examined. By contrast, OpenAI's models exhibited a stronger perceived left-of-center orientation. Interestingly, xAI's Grok models, despite being explicitly marketed as alternatives to ideologically constrained AI systems, were also perceived by respondents as exhibiting a noticeable liberal bias on many political topics.

Importantly, these findings should not necessarily be interpreted as evidence of deliberate political manipulation. Rather, ideological tendencies may emerge as unintended consequences of training data distributions, human preference annotations, reinforcement learning objectives, and institutional definitions of safety and fairness. Consequently, perceived bias often reflects the cumulative effects of numerous design decisions rather than explicit political objectives.

\subsection{Political Regimes and Regulatory Influence}

The nature of LLM censorship also varies systematically across political regimes. In democratic societies, content moderation is generally justified in terms of preventing harm, protecting vulnerable populations, and reducing the dissemination of dangerous information. Typical restrictions target areas such as hate speech, self-harm promotion, terrorism, child exploitation, and illegal activities.

In more authoritarian political systems, censorship objectives frequently extend beyond harm reduction to include the protection of state interests and ideological conformity. Under such conditions, LLMs may actively suppress criticism of government institutions, politically sensitive historical events, opposition movements, or alternative political viewpoints.

China provides one of the most prominent examples of this model of governance. Draft measures issued by the Cyberspace Administration of China in 2023 explicitly place responsibility for model outputs on service providers and require generated content to conform to state-defined standards of legality and social responsibility \cite{Toner2023DigiChina}. Such regulations create strong incentives for aggressive filtering and extensive moderation infrastructures.

Comparable dynamics may emerge in other jurisdictions where governments exert significant control over information flows. Although implementation details vary, the underlying principle remains similar: AI systems are increasingly expected to reflect and enforce local legal, political, and cultural norms.

\subsection{Implications for Global AI Governance}

The observed variation in censorship practices highlights a fundamental challenge for global AI governance. LLMs are often presented as universal technologies; however, their behavior is deeply shaped by local regulatory environments, cultural expectations, and institutional priorities. Different models may therefore provide substantially different responses to identical queries depending on where they were developed, how they were aligned, and which legal frameworks govern their deployment \cite{NIST2023AIRMF,UKDSIT2023AIRegulation}.

\section{Governance, Regulatory, and Legal Considerations}\label{sec:governance}

The rapid deployment of LLMs has prompted increasing regulatory scrutiny concerning content moderation, transparency, accountability, and algorithmic governance. Although comprehensive legal frameworks remain under development in many jurisdictions, policymakers have begun to establish guidelines and statutory requirements that directly or indirectly shape the moderation practices of LLM providers.

\paragraph{United States.}
As of mid-2026, the United States has not enacted legislation specifically governing content moderation by LLMs. The \emph{Executive Order on Safe, Secure, and Trustworthy Artificial Intelligence}, issued on October 30, 2023, encouraged developers to conduct red-teaming exercises, safety evaluations, and risk assessments, but it was rescinded on January 20, 2025 \cite{WhiteHouse2023EO,NIST2025EO}. Consequently, LLM services continue to operate largely under existing legal frameworks applicable to online platforms and under evolving voluntary or agency-specific risk-management practices \cite{NIST2023AIRMF}.

Current U.S. law generally treats LLM providers as private actors exercising editorial discretion. Protections afforded by Section 230 of the Communications Decency Act continue to shield many AI-enabled services from liability associated with user-generated content \cite{CDASection230}. Restrictions on model outputs therefore arise primarily through existing legal doctrines concerning defamation, obscenity, fraud, incitement to violence, and other unlawful forms of expression. Nevertheless, regulatory agencies such as the Federal Trade Commission (FTC) and the Federal Communications Commission (FCC) have increasingly expressed concerns regarding algorithmic transparency, fairness, and consumer protection, suggesting that future regulatory interventions may focus on disclosure obligations and bias mitigation.

\paragraph{European Union.}
The European Union has emerged as a leading jurisdiction in the regulation of artificial intelligence through the enactment of the \emph{Artificial Intelligence Act} (AI Act), adopted in 2024 and scheduled for full implementation beginning in 2026 \cite{EUAIAct2024}. The AI Act establishes a risk-based regulatory framework in which certain applications of generative AI may be classified as high-risk, particularly when deployed in critical sectors or when capable of facilitating deception.

A notable provision of the AI Act is Article~50, which requires providers of certain AI systems to make users aware that they are interacting with an AI system and imposes transparency obligations for synthetic or manipulated content \cite{EUAIAct2024}. Such requirements are intended to enhance transparency and facilitate the identification of AI-generated content. Furthermore, AI systems employed for content moderation must operate consistently with principles established under the \emph{Digital Services Act} (DSA), including fairness, non-discrimination, transparency, and the availability of appeal mechanisms for affected users.

The regulatory framework also mandates extensive documentation, bias assessment, risk management, and post-market monitoring obligations. These requirements may compel providers to systematically evaluate ideological, demographic, and linguistic biases embedded within their models. However, practical implementation challenges remain substantial, particularly for reliable labeling, watermarking, and provenance tracking of non-deterministic AI-generated text.

\paragraph{China.}
China has adopted one of the most comprehensive and restrictive regulatory approaches to generative artificial intelligence. Existing regulations governing ``deep synthesis'' technologies prohibit the generation and dissemination of specified categories of content deemed harmful to state interests, social stability, or public order \cite{CACDeepSynthesis2023}.

Building upon these regulations, the Cyberspace Administration of China (CAC) introduced draft \emph{Measures for Generative Artificial Intelligence Services} in 2023 \cite{Toner2023DigiChina}. These measures establish extensive obligations for model developers and service providers, including requirements that training data be accurate, objective, and consistent with state-approved narratives. Providers may be held legally responsible for content generated by their systems and are required to implement mechanisms that prevent the production of material considered politically sensitive, subversive, or otherwise inconsistent with national policy objectives.

As a result, Chinese LLMs typically incorporate extensive filtering mechanisms throughout the model development pipeline, including dataset curation, alignment procedures, and runtime moderation systems. Violations of content regulations may expose providers to significant administrative sanctions and, in certain circumstances, criminal liability. Consequently, censorship and content control are integrated into the architecture of Chinese LLM services to a much greater extent than in most Western jurisdictions.

\paragraph{Other Jurisdictions.}
Several additional jurisdictions are actively exploring regulatory frameworks for artificial intelligence. The United Kingdom, for example, has pursued a principles-based approach to AI regulation while separately adopting platform-focused online safety legislation \cite{UKDSIT2023AIRegulation,UKOnlineSafetyAct2023}. Current policy discussions generally emphasize transparency, accountability, fairness, and responsible deployment rather than explicit content restrictions. Although many jurisdictions have not yet adopted comprehensive censorship requirements for LLMs, existing laws addressing misinformation, harmful content, and online safety may indirectly influence the operation of generative AI systems.

At the international level, concerns have emerged regarding the growing role of private technology companies as \emph{de facto} arbiters of permissible speech. Free-expression advocates argue that the absence of clear governance frameworks risks transferring significant censorship authority from public institutions to private platform operators. Existing platform regulations, such as Germany's \emph{Netzwerkdurchsetzungsgesetz} (NetzDG), which requires rapid removal of unlawful hate speech, may eventually be extended to AI-mediated content environments \cite{Gorwa2020AlgorithmicModeration}. Consequently, future regulatory developments are likely to focus on balancing content moderation objectives with principles of transparency, due process, and freedom of expression.

\paragraph{Implications for LLM Governance.}
These regulatory developments reveal a growing convergence around transparency, accountability, and risk management as foundational principles of AI governance. However, substantial divergence remains regarding the extent to which governments should influence the substantive content produced by LLMs. While the European Union emphasizes procedural safeguards and accountability, the United States largely relies on market-driven moderation policies, and China imposes direct state oversight of permissible content. These differing approaches are likely to shape the future evolution of LLM censorship, alignment, and content moderation practices across jurisdictions.

\section{Methods for Detecting and Auditing LLM Censorship}
\label{sec:audit}

The increasing role of LLMs as intermediaries of information has motivated the development of systematic methodologies for detecting, measuring, and auditing censorship behaviors. Because modern LLMs are typically proprietary systems with limited transparency regarding their training data, alignment procedures, and moderation policies, researchers must often infer censorship through empirical observation and comparative analysis. Existing approaches can be broadly categorized into behavioral testing, cross-linguistic auditing, model introspection, adversarial evaluation, content-classification analysis, and comparative benchmarking \cite{Liang2023Helm,Raji2020,Madaio2022}.

One of the most widely adopted approaches is \emph{behavioral testing}, in which researchers construct collections of politically, socially, or ethically sensitive prompts and systematically evaluate model responses. By measuring refusal rates, omissions, evasive answers, and framing differences across topics, researchers can identify patterns of content moderation and censorship. For example, longitudinal auditing frameworks such as \emph{AI Watchman} continuously probe commercial LLMs using standardized datasets and track changes in moderation behavior over time. In its summer 2025 audit of social-issue prompts, AI Watchman reported overall refusal rates ranging from approximately 1.2\% for GPT-5 to 3.9\% for GPT-4.1, with DeepSeek refusing approximately 2.5--2.7\% of prompts depending on language \cite{Dai2025Watchman}. These aggregate rates were low for neutral encyclopedic content, but topic-specific rates were much higher: DeepSeek refused Chinese-sensitive Wikipedia content at an average rate of about 31\%, compared with roughly 0.4--5.2\% for OpenAI systems \cite{Dai2025Watchman}. Such approaches are particularly valuable for identifying policy shifts, evolving safety constraints, and inconsistencies in model responses.

Refusal-rate analysis must be interpreted carefully. OpenAI's GPT-5 system documentation describes a shift from hard refusals toward \emph{safe completions}, in which the model attempts to provide helpful high-level information while withholding operationally harmful details \cite{OpenAI2025SafeCompletions}. As a result, a lower hard-refusal rate does not necessarily imply weaker safety controls; it may indicate that the model has shifted from binary denial toward partial, constrained, or non-actionable responses. Longitudinal audits are therefore most informative when refusal rates are paired with qualitative coding of response types, including explicit denials, generic errors, policy explanations, partial answers, and evasive summaries.

A complementary methodology involves \emph{cross-language and cross-regional probing}. Because moderation policies and training corpora often vary across languages and geographic regions, presenting semantically equivalent prompts in different languages can reveal hidden censorship biases. Studies comparing Simplified Chinese and Traditional Chinese prompts have demonstrated systematic differences in model responses, with the former often exhibiting greater caution, neutrality, or evasiveness on politically sensitive topics. Similar comparisons between Chinese and English prompts enable researchers to distinguish censorship effects arising from provider policies from those attributable to technical limitations or linguistic capabilities. This approach is especially important because NLP systems often perform unevenly across languages and social groups, with underrepresented languages receiving less evaluation coverage and weaker safety guarantees \cite{Blodgett2020Bias,Joshi2020LinguisticDiversity}.

Beyond behavioral observation, researchers have begun investigating censorship through \emph{ablation studies and model introspection}. Recent work on hidden factual knowledge suggests that LLMs frequently encode information internally that they do not reveal in generated responses \cite{Gekhman2025}. By probing latent representations, attention patterns, or internal activations, researchers can estimate the extent to which factual knowledge exists within a model but is suppressed by alignment or moderation layers. Such studies provide evidence that censorship may operate not only through the absence of knowledge in training data but also through output-generation constraints that prevent certain information from being expressed.

Another important line of inquiry involves \emph{adversarial testing and red teaming}. Borrowing methodologies from cybersecurity, researchers and practitioners design prompts intended to circumvent moderation safeguards and elicit restricted content. These so-called \emph{jailbreak attacks} serve as stress tests for alignment mechanisms and reveal weaknesses in content-filtering systems \cite{Perez2022RedTeaming,Wei2023Jailbroken,Zou2023Universal}. Internal red teams employed by model developers, as well as independent academic researchers, routinely investigate whether safety controls can be bypassed through prompt engineering, role-playing scenarios, or indirect questioning strategies. Strategic-security research on cyber deception emphasizes that adversarial evaluation should model not only technical exploits but also the beliefs, incentives, and adaptive behavior of attackers and defenders \cite{Zhu2019CyberDeceptionTutorial,PawlickColbertZhu2019Deception}. The success or failure of such attacks provides valuable insight into the robustness of censorship mechanisms.

Researchers have also developed \emph{content-classification auditing frameworks} that attempt to automatically identify censored outputs. These approaches typically involve training classifiers to distinguish between uncensored and potentially moderated responses based on linguistic characteristics such as vagueness, omission patterns, hedging language, or topic avoidance. For example, specialized detection systems have been proposed to identify state-induced censorship biases by analyzing textual features associated with politically sanitized responses, while toxicity benchmarks illustrate how moderation-relevant classifiers can be evaluated at scale \cite{Ahmed2025,Gehman2020RealToxicity}. Such methods enable large-scale quantitative evaluation of censorship behavior across models and domains.

Finally, \emph{comparative benchmarking using open-source models} provides a practical baseline for evaluating moderation practices. Because open-source models often operate with fewer alignment constraints, researchers can compare their responses with those of heavily moderated commercial systems. When an open-source model provides a detailed answer to a query that a proprietary model refuses to address, the discrepancy may indicate the presence of explicit moderation policies rather than technical incapacity. Although such comparisons cannot definitively establish censorship, they offer a useful empirical framework for identifying moderation-related differences in model behavior \cite{Liang2023Helm}.

Audits should also account for temporal instability and stochastic inconsistency. AI Watchman found that some models answered and refused the same Wikipedia content across different collection periods, with GPT-4.1 exhibiting the highest inconsistency rate among the evaluated systems \cite{Dai2025Watchman}. Repeated sampling is therefore necessary for prompts near moderation boundaries, because a single model response may not represent the behavior users would encounter across time, sessions, or deployment updates.

These methodologies demonstrate that effective auditing of LLM censorship requires a multi-faceted approach. Behavioral testing reveals observable moderation outcomes, cross-linguistic analyses expose cultural and regional biases, introspective methods uncover hidden knowledge suppression, adversarial evaluations assess robustness, and comparative benchmarks provide reference points for interpretation. As LLMs continue to play an increasingly influential role in information ecosystems, the development of rigorous, transparent, and reproducible auditing frameworks will be essential for ensuring accountability and understanding the societal consequences of AI-mediated content moderation.

\section{Governance, Transparency, and Mitigation Strategies for LLM Censorship}
\label{sec:mitigation}

The increasing reliance on LLMs for information retrieval, content generation, and decision support has intensified concerns regarding the opacity of content moderation and censorship mechanisms. Because moderation decisions are often embedded within complex pipelines involving training-data curation, alignment procedures, and post-processing filters, researchers, policymakers, and civil society organizations have proposed a variety of governance mechanisms to improve transparency, accountability, and public trust. While no single intervention can fully eliminate the risks associated with algorithmic censorship, a combination of technical transparency, institutional oversight, and regulatory safeguards offers a promising framework for mitigating these concerns \cite{Bommasani2021Foundation,Weidinger2022,NIST2023AIRMF}.

A frequently proposed mechanism is the implementation of \emph{auditable logging systems}. Similar to transparency practices employed in traditional content moderation platforms, LLM providers could maintain secure records of moderation decisions, including refusals, content filtering actions, and the rationale underlying such decisions. Subject to appropriate privacy safeguards, these records could be made available to certified auditors to verify whether specific categories of queries are systematically restricted or treated inconsistently across users. More advanced proposals advocate the creation of \emph{provenance logs} that document the datasets, alignment objectives, rules, and moderation policies influencing individual responses. Such mechanisms would facilitate the detection of hidden biases, policy drift, and unintended moderation outcomes while enhancing institutional accountability \cite{Raji2020,Madaio2022}. They also support trust calibration: users and auditors need enough evidence to determine when an AI system is reliable, when it is withholding information, and when its governance constraints are changing over time \cite{GeZhu2024TrustAI}.

Transparency can be further enhanced through comprehensive \emph{model cards}, \emph{system cards}, and related documentation. Major AI developers have already begun publishing reports describing model architectures, training procedures, safety mechanisms, capabilities, and known limitations. However, these disclosures often provide only high-level summaries of alignment objectives and moderation policies. More detailed documentation could explicitly identify prohibited content categories, multilingual limitations, known failure modes, refusal behaviors, and the rationale behind moderation decisions. Such disclosures would enable researchers, regulators, and users to better understand how content moderation is implemented and where potential biases may emerge \cite{Mitchell2019ModelCards,Gebru2021Datasheets,OpenAI2024RedTeaming,AnthropicConstitution}.

Another important proposal concerns \emph{contestable moderation}. Rather than treating moderation decisions as final and opaque, users could be provided with mechanisms to challenge or appeal refusals. For example, systems might explain the basis of a refusal, allow users to reformulate queries, or provide structured review processes for disputed decisions. More sophisticated implementations could introduce tiered moderation frameworks in which certain restrictions may be overridden under carefully controlled conditions, with all exceptions recorded for subsequent auditing. Although such approaches raise concerns regarding misuse and regulatory compliance, they may reduce the concentration of informational authority within AI systems while generating valuable feedback for refining moderation policies. Ethical frameworks developed for defensive cyber deception similarly emphasize that protective interventions should be constrained by transparency, proportionality, no-harm principles, and fairness, principles that are directly relevant to the design of contestable AI moderation systems \cite{Zhu2023Doctrine,NIST2023AIRMF,UKDSIT2023AIRegulation}.

Closely related is the concept of \emph{pluralistic alignment}, which recognizes that content moderation inevitably reflects normative judgments about acceptable speech and behavior. Rather than aligning models to a single value system, pluralistic alignment seeks to incorporate diverse stakeholder perspectives into governance processes. This may involve configurable moderation settings, multiple alignment profiles reflecting distinct cultural or political viewpoints, or oversight boards composed of representatives from diverse communities. Proponents argue that openly deliberated standards are preferable to hidden alignment objectives because they render value judgments transparent and subject to public scrutiny. At the same time, excessive personalization of moderation policies may contribute to ideological fragmentation and informational echo chambers \cite{Hall2025Slant,ECNL2025,Sorensen2024PluralisticAlignment}.

Regulatory oversight represents another increasingly important mechanism for addressing concerns regarding opaque censorship. Governments around the world are beginning to establish legal frameworks governing transparency, accountability, and risk management in AI systems. The European Union's Artificial Intelligence Act introduces requirements for transparency, documentation, post-deployment monitoring, and risk assessment for high-risk AI systems. Similarly, the Digital Services Act establishes obligations related to content moderation, fairness, and accountability for online platforms. In contrast, some jurisdictions impose substantive restrictions on permissible content, directly influencing the outputs of deployed AI systems. While regulatory intervention can establish minimum standards for transparency and accountability, it may also increase the risk of formalized censorship if content restrictions become overly prescriptive \cite{EUAIAct2024,DigitalServicesAct2022,WhiteHouse2023EO,Toner2023DigiChina}.

Technical approaches to transparency further include \emph{provenance tracking} and \emph{watermarking}. Emerging regulatory proposals envision mechanisms that identify AI-generated content and document its origins. In principle, such systems could allow users and auditors to determine whether content has been modified, filtered, or generated under specific moderation constraints. However, robust watermarking of generated text remains an open research challenge, and current techniques offer only limited guarantees. Consequently, provenance systems should be viewed as complementary tools rather than complete solutions to transparency concerns.

Researchers have also emphasized the importance of \emph{third-party auditing and independent oversight}. External auditing organizations, academic researchers, and civil society groups can provide assessments of moderation behavior that are independent of platform operators. Public benchmarking initiatives, transparency reports, and standardized evaluation frameworks create incentives for providers to maintain consistent moderation practices while enabling cross-model comparisons. Legislative proposals in several jurisdictions have similarly explored requirements for sharing information with accredited auditors as a means of improving accountability \cite{Raji2020,Madaio2022}.

A final avenue for improving transparency involves the development of \emph{open-source models} and \emph{community-governed models}. Open-source LLMs enable researchers to inspect training procedures, alignment methods, and moderation mechanisms directly, thereby reducing the opacity associated with proprietary systems. Community-led models may additionally incorporate local knowledge, cultural diversity, and rights-based governance principles that are difficult to achieve within centralized corporate platforms. By allowing public scrutiny of moderation rules and implementation choices, such models can serve as valuable counterweights to highly centralized AI ecosystems \cite{ECNL2025,Bommasani2021Foundation}.

Ultimately, addressing concerns surrounding LLM censorship requires more than technical intervention alone. Effective governance will likely depend upon the integration of transparency mechanisms, independent auditing, user participation, regulatory oversight, and ongoing public deliberation regarding acceptable moderation practices. User feedback systems, appeal processes, and multidisciplinary governance frameworks can help identify instances of over-censorship while preserving protections against genuinely harmful content. The central challenge is therefore not merely to reduce censorship, but to ensure that moderation decisions are transparent, contestable, and accountable to the communities they affect. In this context, transparency emerges as a foundational principle: users should not only know that content has been restricted, but also understand why such restrictions have been imposed and how those decisions can be challenged or reviewed. This points toward a resilience-oriented governance model in which moderation systems are continuously monitored, stress-tested, updated, and recovered from failures rather than treated as one-time compliance artifacts \cite{Zhu2025CyberResilience}.

\section{Societal Implications of LLM Content Moderation and Censorship}\label{sec:societal}

The increasing adoption of LLMs as interfaces for information retrieval, search, education, and decision support has elevated their role from passive information-processing systems to active gatekeepers of knowledge. Consequently, content moderation and censorship mechanisms embedded within LLMs have significant implications for how individuals access information, form opinions, and participate in public discourse. Unlike traditional search engines, which typically present users with multiple sources and competing viewpoints, LLMs generate a single synthesized response. This structural characteristic amplifies the influence of content moderation policies because any omission, filtering decision, or framing choice directly shapes the information environment experienced by the user \cite{Gorwa2020AlgorithmicModeration,Bender2021StochasticParrots}.

One important consequence is the emergence of what may be termed \emph{algorithmic gatekeeping}. As users increasingly rely on conversational AI systems rather than conventional search engines, the model effectively determines which information is presented and which information remains inaccessible. Civil society organizations have warned that a small number of dominant foundation models may establish an \emph{algorithmic monoculture} in which a limited set of providers implicitly defines global norms regarding acceptable speech and information dissemination \cite{ECNL2025}. Under such conditions, content moderation decisions are no longer merely technical safeguards but become mechanisms that shape public knowledge and discourse.

A related concern arises from the tendency of LLMs to provide a single authoritative answer rather than a collection of alternative sources. Although this approach enhances convenience and usability, it also concentrates informational power within the model itself. Users may be unaware of omitted perspectives, competing interpretations, or alternative factual accounts. Consequently, soft censorship---manifested through omissions, selective emphasis, or framing effects---can be particularly difficult to detect. Even when a model does not explicitly refuse to answer a query, its response may subtly privilege certain viewpoints while marginalizing others, thereby influencing the user's understanding of a topic without transparent disclosure.

Beyond information access, LLMs possess significant persuasive capabilities. Recent empirical studies demonstrate that LLM-generated arguments can influence attitudes and policy preferences at levels comparable to traditional human-generated persuasion campaigns \cite{Bai2025Persuasion}. Exposure to carefully constructed AI-generated messages has been shown to shift individuals' positions on public policy issues, even when participants are informed that the content was produced by an artificial intelligence system. Systematic omissions or consistent framing biases may therefore gradually shape public opinion and political attitudes over time.

Content moderation practices also influence public trust in AI systems. While restrictions on harmful content are generally intended to reduce risks associated with misinformation, hate speech, violence, and self-harm, excessive filtering may produce unintended consequences. If users perceive that information is being selectively withheld or ideologically filtered, confidence in AI-generated outputs may diminish. Conversely, insufficient moderation can expose users to harmful, misleading, or dangerous content. This tension highlights a fundamental governance challenge: balancing openness and informational completeness against legitimate concerns regarding safety and societal harm.

The psychosocial effects of content moderation further complicate this balance. Restrictions on content related to self-harm, extremism, hate speech, or dangerous activities may protect vulnerable individuals and reduce opportunities for malicious exploitation. However, overly restrictive moderation systems may inadvertently suppress legitimate discussions concerning mental health, political controversy, historical events, or social identity. Such over-enforcement risks chilling lawful expression and disproportionately affecting marginalized communities whose experiences, linguistic patterns, or cultural perspectives may be misclassified as problematic by moderation algorithms trained primarily on dominant-language datasets \cite{ECNL2025,Blodgett2020Bias,Joshi2020LinguisticDiversity}.

As LLMs increasingly function as primary intermediaries between users and knowledge, their moderation policies influence information diversity, public trust, civic discourse, and the formation of beliefs. The central challenge for researchers, policymakers, and developers is to design moderation frameworks that effectively mitigate harm while preserving transparency, pluralism, and freedom of expression in increasingly AI-mediated information environments.

A central governance difficulty is the trade-off between over-filtering and under-filtering. Over-filtering occurs when systems refuse or suppress benign, educational, journalistic, artistic, or minority-community speech; under-filtering occurs when systems allow genuinely harmful content to pass through. The European Center for Not-for-Profit Law warns that LLM-based moderation systems can over- or under-enforce content policies and that such errors may disproportionately affect marginalized communities, particularly in non-dominant languages and Global Majority contexts \cite{ECNL2025}. These risks reinforce the need for moderation systems that expose uncertainty, permit contestation, and are evaluated not only for average accuracy but also for disparate impact across languages, communities, and political contexts.

\section{Open Challenges and Future Research Directions}\label{sec:open}

Despite growing scholarly and regulatory attention to content moderation in LLMs, several conceptual, empirical, and governance challenges remain unresolved. A central difficulty concerns the scope of the term \emph{censorship}. Alignment-based filtering is often defended by providers as a necessary mechanism for harm reduction, particularly in domains involving violence, self-harm, hate speech, fraud, or other unlawful activity. However, such defenses rely on contested assumptions about what counts as ``harmful'' content and who has the authority to define it. If a model's operationalization of safety excludes legitimate political, cultural, historical, or minority perspectives, then safety-oriented filtering may function as a form of epistemic suppression. The boundary between responsible moderation and censorship therefore remains analytically unstable \cite{Casper2023RLHFLimitations,Sorensen2024PluralisticAlignment}.

A related challenge concerns the trade-off between openness and safety. Overly permissive systems may enable harmful behavior, while overly restrictive systems may become unreliable or unusable for legitimate inquiry. For example, excessive filtering may suppress benign discussion of sensitive topics, including mental health, political conflict, sexuality, or LGBTQ-related issues, merely because these domains are associated with possible abuse. The central research problem is therefore not whether LLMs should moderate content at all, but how moderation systems can minimize harmful outputs without suppressing lawful, educational, journalistic, or socially valuable speech.

Regulatory uncertainty further complicates this problem. Although emerging legal frameworks such as the European Union's Artificial Intelligence Act establish transparency, documentation, and risk-management obligations, many implementation questions remain unresolved. It is unclear how requirements for disclosure, watermarking, provenance, post-market monitoring, and cross-platform transparency can be technically implemented for non-deterministic generative models. Moreover, global enforcement remains difficult because LLM providers operate across jurisdictions with conflicting legal norms. As a result, regulatory frameworks may improve accountability in some contexts while simultaneously creating incentives for more formalized or jurisdiction-specific censorship.

The problem of global norms is equally unresolved. LLMs are deployed across societies with different legal systems, political expectations, cultural values, and speech traditions. A model optimized for one jurisdiction may be viewed as irresponsible in another or censorial in a third. For example, a Western user may expect unrestricted discussion of democracy, civil liberties, or state repression, whereas a model deployed under authoritarian regulatory constraints may be legally required to suppress or sanitize such content. One proposed response is to promote ideological and geographic diversity among models, allowing users to select systems that better reflect their informational and cultural expectations. However, this approach also risks reinforcing ideological self-selection, filter bubbles, and epistemic fragmentation if users choose only models that confirm their prior beliefs.

New attack vectors are also likely to emerge as LLMs become embedded in search engines, messaging systems, productivity tools, social platforms, and enterprise workflows. In such settings, censorship may occur not only at the model-output layer but also at the platform, application, retrieval, ranking, or user-interface layer. For instance, a messaging application that uses an LLM to screen or rewrite user-to-user communication may introduce indirect forms of moderation that are difficult to observe through conventional model audits. These forms of mediated censorship may evade existing accountability mechanisms because the moderation decision is distributed across multiple technical components rather than located within a single model response. The rise of agentic AI further complicates this problem: LLM-powered agents can plan, invoke tools, coordinate across workflows, and act on external systems, making moderation a property of multi-agent execution environments rather than isolated text generation \cite{Zhu2025GameTheoryLLM,YangZhu2026InternetAgenticAI}.

Several research gaps follow from these unresolved issues. First, richer multilingual and cross-cultural audits are needed. Existing studies disproportionately focus on English and Chinese, leaving comparatively little evidence regarding censorship behavior in languages such as Arabic, Hindi, Swahili, Russian, Persian, or Indigenous languages. Broader multilingual testing is essential for identifying whether moderation systems disproportionately suppress underrepresented linguistic communities or culturally specific forms of expression \cite{Blodgett2020Bias,Joshi2020LinguisticDiversity}.

Second, the field lacks standardized metrics for measuring \emph{soft censorship}. Explicit refusals can be counted through refusal-rate analysis, but omissions, evasive framing, selective emphasis, and strategic ambiguity are more difficult to quantify \cite{Noels2025,Ahmed2025}. Future benchmarks should evaluate not only whether a model answers a query, but also whether the answer is complete, balanced, contextually faithful, and representative of relevant viewpoints. This may require hybrid evaluation methods combining human expert review, factual databases, retrieval-based comparison, and cross-model disagreement analysis.

Third, more empirical work is needed on user-level impacts. It remains unclear how repeated exposure to filtered or one-sided LLM responses affects knowledge formation, political attitudes, trust in institutions, or perceptions of contested social issues. Longitudinal studies could examine whether users become more confident in incomplete information, whether censored responses increase distrust in AI systems, and whether model-mediated framing contributes to polarization or the marginalization of minority viewpoints.

Fourth, methods for probing hidden or suppressed knowledge require further development. If LLMs internally encode factual information that they do not express in ordinary outputs, researchers need techniques for distinguishing between genuine knowledge absence and alignment-induced suppression \cite{Gekhman2025}. Indirect prompting, representation probing, layer-wise analysis, and controlled ablation studies may help determine whether censorship arises from training-data omissions, learned refusal behavior, post-processing filters, or higher-level alignment objectives.

Fifth, dynamic adaptation experiments are needed because moderation policies evolve over time. Providers frequently update system prompts, safety classifiers, refusal policies, and alignment objectives, making censorship behavior a moving target. Controlled experiments that compare model behavior before and after policy changes could clarify how small adjustments in safety settings affect refusal rates, answer completeness, ideological framing, and user trust. Such studies would make the safety--censorship trade-off more empirically tractable \cite{Dai2025Watchman}.

Sixth, governance prototypes should be developed and evaluated in practice. Examples include contestable-AI interfaces that provide explanations for refusals, appeal mechanisms for disputed moderation outcomes, public dashboards reporting refusal rates by topic, and configurable moderation settings with transparent constraints. These prototypes should be tested not only for usability but also for abuse resistance, fairness, and their effect on public trust \cite{NIST2023AIRMF,UKDSIT2023AIRegulation}. Economic and institutional mechanisms also deserve attention: contract-theoretic pricing, liability allocation, and insurance frameworks for agentic AI suggest possible ways to connect transparency, service quality, and risk-bearing responsibilities in deployed AI systems \cite{YangZhu2025PACT,Zhu2026InsuranceAgenticAI}.

Finally, the research community would benefit from shared datasets, open benchmarks, and collaborative audit infrastructures. Publicly available test suites covering politically sensitive questions, hate-speech edge cases, multilingual prompts, historical controversies, and safety-relevant domains would allow more consistent comparison across providers \cite{Liang2023Helm,Dodge2021C4}. Longitudinal initiatives such as AI Watchman could be expanded to include more models, languages, jurisdictions, and content categories, with data made available for independent replication.

\section{Conclusion}\label{sec:conclusion}
LLM censorship is neither a purely technical artifact nor a simple policy choice. It emerges from the interaction of training data, alignment objectives, legal constraints, corporate norms, platform incentives, and user expectations. Because LLMs increasingly act as interfaces to knowledge, their moderation decisions can shape not only what users are able to see, but also how contested issues are framed, prioritized, or silently omitted.

Some degree of content moderation may be unavoidable in widely deployed AI systems, particularly where models could enable violence, fraud, abuse, or other harms. However, the legitimacy of such moderation depends on whether it is transparent, proportionate, contestable, and accountable. Future governance should therefore move beyond the binary question of whether LLMs censor content and instead ask how moderation decisions are made, who controls them, whose values they encode, and how affected users can inspect, challenge, or diversify them. The long-term objective should be a pluralistic and auditable AI ecosystem in which safety protections do not become instruments of opaque epistemic control.

\bibliographystyle{unsrt}
\bibliography{refs}

\end{document}